\documentclass[reprint,pre,superscriptaddress]{revtex4-1}
\usepackage{amsmath}
\usepackage{bm}
\usepackage{graphicx,latexsym}
\usepackage{verbatim}
\usepackage{enumerate}
\usepackage{ulem}
\usepackage[table]{xcolor}
\usepackage[colorlinks,allcolors=.,citecolor=blue]{hyperref}

\newcommand{\MSD}{\langle d^2(t)\rangle}
\newcommand{\figformat}[1]{{#1}}

\begin{document}

\title{Enhanced flagellar transport in asymmetric periodic arrays}
\author{James L. Kingsley}
\affiliation{Department of Physics, Worcester Polytechnic Institute, Worcester, MA 01609}
\author{Utkan Demirci}
\thanks{co-corresponding authors}
\affiliation{Bio-Acoustic MEMS in Medicine BAMM Laboratory, Canary Center at Stanford for Cancer Early Detection, Department of Radiology, Stanford School of Medicine,
Palo Alto, CA 94304}
\affiliation{Department of Electrical Engineering (by courtesy), Stanford University, Stanford, CA, USA94304}
\author{Erkan T\"{u}zel}
\thanks{co-corresponding authors}

\affiliation{Department of Physics, Worcester Polytechnic Institute, Worcester, MA 01609}

\date{\today}

\begin{abstract}
Active swimmers are ubiquitous in nature, found in many diverse biological systems ranging from bacteria to vertebrate fish.
Of particular importance are sperm cells which are swimmers that are crucial for the survival of many species including humans.
Despite decades of work, the fluid physics of sperm in complex micro-environments such as the cervical tract, or the microfluidic devices used in assistive reproductive technologies remains illusive.
Recently, a novel microfluidic device featuring periodic post arrays has been developed, and shown to select sperm cells with better motility, morphology and DNA integrity, more efficiently than existing approaches.
Motivated by this, here we present a multi-scale model that aims to provide insight physical insight into the motility behavior of sperm in such periodic geometries.
Our model combines a fluctuating hydrodynamic model of sperm with a probabilistic discrete-time lattice approach, and we show how hydrodynamic and boundary interactions facilitate both the enhancement of speed and persistence length of sperm cells in this post array.
We then discuss how this enhancement of flagellar transport is related to its propensity, and develop a phase diagram.
Our findings not only shed light into the fluid physics of flagellar swimmers in periodic arrays, but also have direct implications in a broad range of areas beyond fertility, including bio-inspired robotics, disease detection and drug delivery.
\end{abstract}

\maketitle

\newcommand{\mv}[1] {{\bm#1}}

\section{Introduction}

Active swimmers are found at every scale in nature, appearing in many diverse biological systems, ranging from bacteria to sperm cells, in 
multiple environments from marine ecosystems to mammals, and have been an inspiration for the design of artificial swimmers in the past few decades~\cite{Lauga2009}. 
In particular, sperm motility---due to its clinical, biological and technological biological importance---has been the subject of hundreds of experimental and theoretical studies, going back to Lord Rothschild who first observed that spermatozoa preferentially concentrated near boundaries as the result of hydrodynamic interactions\cite{Rothschild1963}.
In fact, sperm, as well as many other active swimmers, have been an inspiration for the design of artificial micron-scale flagellar systems including bio-hybrid microrobots~\cite{Sitti2009,Lum2016,Magdanz2013,MedinaSanchez2015,Magdanz2017}, and have broad applications in minimally invasive surgery techniques\cite{Nelson2010}, bio-sensing~\cite{Guix2014} and targeted drug delivery~\cite{Han2016}. 

With the goal of understanding the complex  interactions between these active swimmers and their environments, many models have been developed, including fully hydrodynamic approaches~\cite{Gaffney2011,Elgeti2015,Yang2008,Fauci1995,Smith2009}.
While considerable progress has been made in understanding the beating patterns, the coupling of hydrodynamics and chemotaxis~\cite{Jikeli2015}, synchronization of sperm flagella~\cite{Mettot2011} and attractive interactions induced by hydrodynamics~\cite{Lauga2006,Elfring2010}, to this date, very few models exist~\cite{Elgeti2015} that can quantitatively model complex geometries present in more realistic scenarios (large time and length scales), such as those found in some of the microfluidic devices used in Assistive Reproductive Technologies (ARTs)
for treatment of infertility~\cite{Agarwal2015,Turchi2015, Tasoglu2013}.

\begin{figure}[htb] 
 \centering
\includegraphics[width=8.6cm]{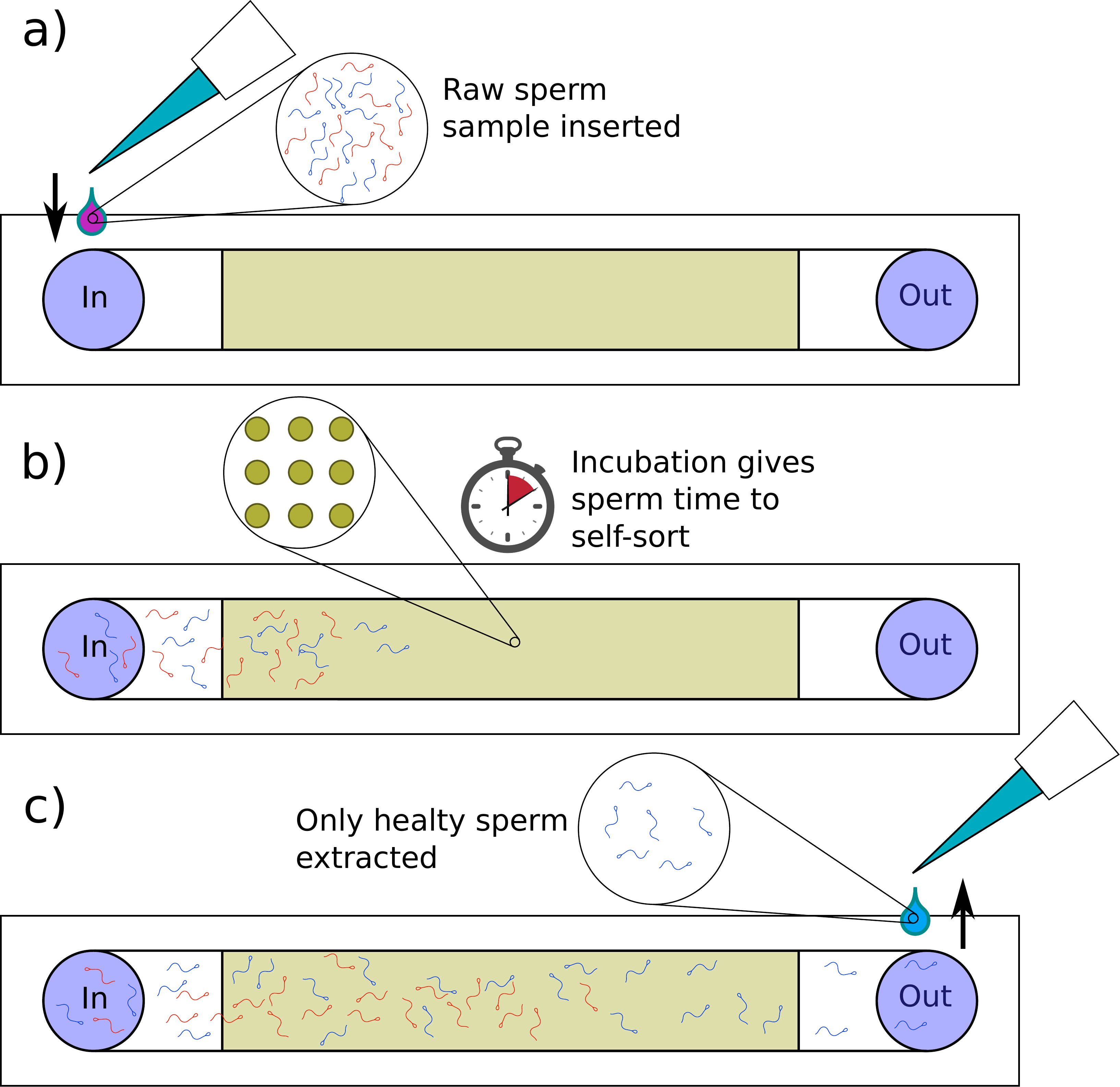}
\caption{Overview of a post-array microfluidic device for sperm sorting.
a) A sample of raw semen, containing a mixture of healthy and unhealthy sperm (as well as other impurities), is placed into the inlet of the device.
b) The device is incubated for some period of time, during which the sperm traverse the length device.
In a well-designed device, sperm with high degree of motility and void of morphological defects are capable of swimming to the outlet in an effective manner. 
c) Finally, sperm are extracted from the outlet for use in IVF applications. See Refs.~\cite{Chinnasamy2017} for details. 
}
\label{fig:design}
\end{figure}

In particular, identification and isolation of the most motile and healthy sperm (with DNA integrity) is essential for 
In Vitro Fertilization (IVF) and Intra-Cytoplasmic Sperm Injection (ICSI)~\cite{Ombelet2008, Tasoglu2013}. 
Inspired by the space-constrained environment of the reproductive tract, in the past decade we~\cite{Agarwal2015,Tasoglu2013} and 
others~\cite{Zhang2015,Nosrati2016} developed microfluidic devices that utilize 
hydrodynamic interactions to address sperm selection challenges. These approaches increased the prevalence of microfluidic devices in clinical settings over traditional techniques such as density gradient centrifugation or swim-up~\cite{Henkel2003,Asghar2014}, further
highlighting the importance of understanding sperm behavior in such complex micro-geometries. 

More recently, a new microfluidic device featuring periodic post arrays (see Fig. \ref{fig:design} for an illustration) has been proposed~\cite{Chinnasamy2017,Demirci2016}, demonstrating an ability to increase the temporal and spatial separation between progressive and non-progressive motile sperm populations, improving the ability to select sperm with normal morphology nearly 5-fold, and reducing the processing time at the clinic by 3-fold~\cite{Chinnasamy2017}, compared to previous methods.

The use of periodic micro-arrays is not uncommon in the microfluidic community, and a number of studies used such symmetric or asymmetric post arrays, ranging from applications to particle sorting and separation~\cite{Morton2008,McGrath2014,Zeming2016,Du2016} to circulating 
tumor cell capture~\cite{Nagrath2007}.
However, none of the existing approaches leverage the intrinsic motiility of active swimmers to self-sort in a periodic array, in the absence of flow.

The ability of sperm to self-sort~\cite{Chinnasamy2017} in this environment with such success is rather surprising; while it is 
expected for hydrodynamics to boost motility in confined geometries, in a flow-free system the posts can also trap the sperm hindering its motility. 

In this paper, we investigate the fluid physics of sperm moving through this post array geometry, and show that the channels act as a transport enhancer if tuned properly.
Organization of the paper is as follows: our modeling approach is described in Sections  \ref{sec:MPCD} and \ref{sec:lattice-model}, 
followed by our simulation results in Section \ref{sec:results}, including a complete phase diagram of sperm transport in post arrays. 
We end the paper with a discussion of our results and outlook for the future. 

\section{Sperm in an MPCD Fluid} \label{sec:MPCD}

In order to model the sperm behavior in the post-array geometries, we use an approach combining the strength of 
fully hydrodynamic particle-based simulations that capture detailed sperm motility, along with the computational efficiency of a probabilistic model (discrete-time lattice model with rotation). More specifically, we start by characterizing the microscopic behavior of sperm in a given post geometry using a bead-and-spring sperm model embedded in a particle-based solvent. 
Once enough trajectory data is collected for a wide range of post spacings, we use sperm motility at this scale to calculate probabilities for a stochastic model that can be used to model sperm motility behavior within a larger post array. One advantage of this multi-scale approach is that it can then be used to optimize sperm sorting devices featuring posts with realistic device dimensions containing large populations of sperm, at a fraction of the computational cost 
(see Ref. \cite{Chinnasamy2017} for details). In what follows, we describe the details of this model. Given the aspect ratio of the microfluidic channels typically used ($< 50~\mu m$) and the sperm tail size, the simulations are restricted to two dimensions in order to reduce computational complexity. All the particle-based simulations (described below) are performed using a GPU implementation with Nvidia CUDA~\cite{Chen2013}. 

\subsection{MPCD Solvent}
\label{sec:solvent-model}
In multi-scale simulations of fluids, the majority of the computational cost is from modeling the surrounding fluid, limiting the system size and the number of embedded structures that can be used. To overcome this limitation, here we use a  particle-based simulation technique, namely Multi-Particle Collision Dynamics (MPCD) (also known as Stochastic Rotation Dynamics, SRD), to model the solvent degrees of freedom \cite{malevanets1999mesoscopic,Ihle2001,Kapral2008,Gompper2009,Messlinger2009}. 
Transport behavior of the MPCD fluid has been very well 
characterized in both two and three dimensions, including its complete description of equilibrium transport properties\citep{Tuzel2003,Tuzel2006,Ripoll2005}.
Over the past decade, this approach has been used extensively to study many diverse problems in the field of complex fluids and polymeric 
mixtures\citep{Ali2004,Tuzel2006,Tucci2004,Winkler2013} including the behavior of vesicles and red blood cells in hydrodynamic flow \citep{Noguchi2004,Noguchi2005,Noguchi2006,LiamMcWhirter2012,Noguchi2010,Noguchi2010a}, chemically-driven 
transport~\cite{Shen2016,Huang2016,Reigh2015,Thakur2012,Thakur2011}, transport in 
crowded environments~\cite{Huang2017}, star polymers~\cite{Ripoll2007},viscoelastic fluids in shear flow\citep{Tao2008}, and swimmers including sperm \citep{Reigh2013,Yang2008,Elgeti2010,Theers2016}, self-phoretic microswimmers~\cite{Yang2016,Yang2014,Costanzo2014}.

\begin{figure}[htb] 
 \centering
\includegraphics[width=8.6cm]{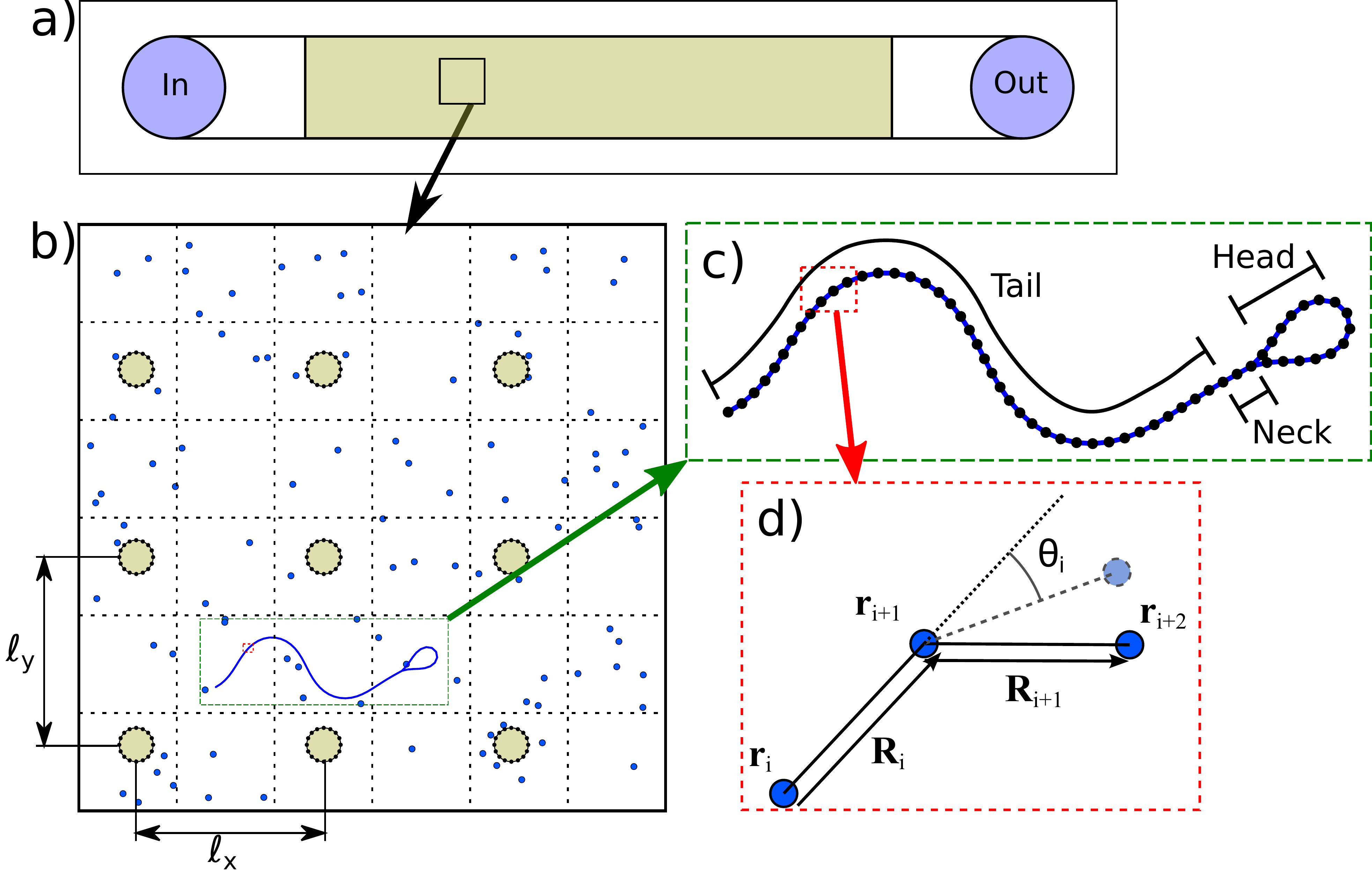}
\caption{Overview of the implementation of the MPCD model.  a) Typical microfluidic device design in which the post array zone is highlighted. 
b) A closer look at the post zone, showing the fluid particles, posts (with spacing $\ell_x$ by $\ell_y$), and sperm.
c) Layout of bead-and-spring model of the sperm, showing the head, neck, and tail regions.
d) Detail of a single trimer bond, indicating point positions ${\bf r}_i$ and tangent vectors ${\bf R}_i$. 
A bend is imposed in the bond between ${\bf R}_i$ and ${\bf R}_{i+1}$, by setting the angle of minimum energy, $\theta_i$.
Diagram is not to scale, as it spans four orders of magnitude of length scales.
}
\label{fig:MPCDModel}
\end{figure}

In the MPCD approach, the fluid is modeled as a set of point particles with identical mass $m$, position $\mv{r}_i$, and velocity $\mv{v}_i$.
The simulation proceeds in discrete time steps of length $\tau$, with each time step consisting of two processes. In the first (streaming) step, the particles move ballistically with
\begin{equation}
\mv{r}_i(t+\tau) = \mv{r}_i(t)+ \mv{v}_i (t) \tau \;\;. 
\end{equation}
In the second (collision) step, the particles exchange momentum with their local neighbors in a single collective collision\citep{malevanets1999mesoscopic}.
A grid of boxes of side length $a$ is imposed over the simulation volume (see Fig. \ref{fig:MPCDModel}). For each collision cell, the total mass and momentum of the particles within that box is summed, yielding an average cell velocity $\mv{u}_{c_i}$.
From this, the collision rule for a particle $i$ in a cell $c_i$ is
\begin{equation}
\mv{v}_i(t+\tau) = \mv{u}_{c_i}+\mathcal{R}\left(\mv{v}_i (t)-\mv{u}_{c_i}(t)\right) \;\;. \label{eq:collision}
\end{equation}
Here, $\mathcal{R}$ represents a rotation by an angle $\pm \alpha$, an open adjustment parameter to adjust fluid properties such as viscosity\citep{Tuzel2006}, whose direction is chosen randomly for each collision cell.
A random-shift of the collision grid in the range $[-a/2,a/2]$ before the collision step guarantees Galilean invariance\citep{Ihle2001}. 

The shear viscosity of the MPCD solvent in two dimensions is given by\citep{Tuzel2006, Kapral2008,Gompper2009}
\begin{eqnarray}
\hspace{-1cm} \nu = \frac{k_B T \tau}{2m} && \left(\frac{\rho}{(\rho-1+e^{-\rho})\sin^2 \alpha} -1 \right) +  \nonumber \\
       && +\frac{a^2}{12\tau} (\rho-1+ e^{-\rho}) \left( 1- \cos \alpha\right)  \;\;,
\end{eqnarray}
\noindent where $\rho$ is the particle density, $k_B$ is Boltzmann's constant, and $T$ is the temperature. 
Parameters used in the simulations (see  Table \ref{tab:MPCD_Parameters}) yield a viscosity value of $\nu \simeq 3.0$ in units of $a^2/\tau$, which combined with an upper limit of observed sperm speed of approximately $0.025$ (in units of $a/\tau$), and a sperm length of $\sim 60$ (in units of $a$), gives a Reynolds number of $Re \le 0.5$.
If we consider the post spacing to be the relevant length-scale, one would get $Re \le 0.25$ instead, for the upper bound.  These  bounds for the Reynolds number are consistent with the earlier 
sperm modeling studies that employed MPCD~\cite{Yang2008,Elgeti2010}, and ensure the simulations are done in a regime where inertia does not play a significant role. 

\begin{table}[ht] 
 \centering
\begin{ruledtabular}
 \begin{tabular}{lrlrlr}
 \multicolumn{2}{c}{\textbf{Solvent}}     &     \multicolumn{2}{c}{\textbf{Sperm}}         &     \multicolumn{2}{c}{\textbf{Post}}         \\ \hline
$L_x$                     & $6 \ell_x$             &     $k_\text{s}$             & $10^6 k_BT/a^2$    &     $\delta$     & $a$              \\
$L_y$                     & $6 \ell_y$              &     $\kappa_\text{b}$    & $3000 k_BT$          &     $m_p$             & $5$              \\
$a$                     & 1.0                      &     $A$                          & 1.0                          &     $r_\text{p}$               & $3.6a$            \\
$k_BT$               & 1.0                      &     $\omega$                & $0.0005\pi/\tau$     &     $\ell_x$                     & $[16-48]a$                \\
$\tau$                  & 0.025                 &     $k$                           & $0.2/a$                   &     $\ell_y$                     & $[16-48]a$         \\
$\tau_\text{MD}$ & $\tau/100$         &     $m_s$                      & $10 m$                   &      $\epsilon$  & $50 k_BT$             \\
$\alpha$              & $\pi/2$               &     $d_h$                       & $5  a$                     &      $\sigma$    & $a$        \\
$m$                     & 1.0                    &     $\ell_\text{neck}$       & $4  a$                    &     & \\
$\rho$                  & $10/a^2$          &     $\ell_\text{tail}$          & $50 a$                   &     & \\
 \end{tabular}
 \caption{Parameters used in the MPCD simulations. Here, $d_h$  denotes sperm head diameter, $\ell_\text{neck}$ and $\ell_\text{tail}$ correspond to the sperm neck and tail length in lattice units, respectively.  $r_\text{p}$ denotes the post radius. The remaining parameters are described in the text.  }
 \label{tab:MPCD_Parameters}
 \end{ruledtabular}
\end{table}

\subsection{Bead-and-Spring Sperm Model}
\label{sec:sperm-model}
We model the individual sperm cells as bead-spring chains embedded in the MPCD solvent (see Fig. \ref{fig:MPCDModel}), subject to the following potential 
\begin{eqnarray}
 U &=& \frac{k_\text{s}}{2} \sum_{i=0}^{n-1}(|\mv{R}_{i}|-\ell)^2
      +\frac{\kappa_\text{b}}{2} \sum_{i=0}^{n-2}\left(\frac{\mv{R}_{i}}{|\mv{R}_i|}-\mathcal{R}_i\frac{\mv{R}_{i+1}}{|\mv{R}_{i+1}|}\right)^2 \nonumber  \\
 &=& \frac{k_\text{s}}{2}\sum_{i=0}^{n-1}(|\mv{R}_{i}|-\ell)^2
      +\kappa_{\text{b}}\sum_{i=0}^{n-2}\frac{\mv{R}_{i}\cdot\mathcal{R}_i\mv{R}_{i+1}}{|\mv{R}_{i}||\mv{R}_{i+1}|}  \;\;. 
\end{eqnarray}
\noindent as employed in earlier studies~\cite{Yang2008}. Here, the constants $k_\text{s}$ and $\kappa_\text{b}$ correspond to the stretching and bending stiffness values, respectively, and the values used in the simulations are given in Table \ref{tab:MPCD_Parameters}. Each bond in the polymer backbone is defined as
$ \mv{R}_{i} \equiv \mv{r}_{i+1}-\mv{r}_{i}$, where $\mv{r}_i$ is the position of the bead $i$, and $\ell$ is the equilibrium bond length. Finally, 
non-straight bonds (when there is an intrinsic bend imposed by a traveling wave) are accomplished by rotating the second of the segments making up the $i^{th}$ bond by $\theta_i$, the angle of minimum energy for that bond, using the rotation matrix
\begin{equation}
 \mathcal{R}_i = \left(\begin{array}{cc}
		\cos\theta_i & \sin\theta_i \\
		- \sin\theta_i & \cos\theta_i \\
             \end{array}\right) \;\;.
\end{equation} 
For the head and neck of sperm, $\theta_i=0$, while for the tail we impose a traveling wave given by
\begin{equation}
 \theta_i(t) = A\sin(ki+\omega t+\phi_0) \;\;,
\end{equation}
\noindent where $A$ is the wave amplitude, $k$ is the wave number, $\omega$ is the angular frequency of sperm tail oscillation. The phase angle $\phi$ is set to zero. Once again all the relevant simulation parameters are given in Table \ref{tab:MPCD_Parameters}. The sperm parameters are chosen to mimic human sperm motility in typical microfluidic device conditions~\cite{Chinnasamy2017,Tasoglu2013,Asghar2014}. 

\subsection{Boundary Conditions and Solvent Coupling}

The solvent particles and sperm are constrained within the simulation area by imposing periodic boundary conditions.
This condition simply acts by enforcing that $L r_i < \lfloor L r_i\rfloor$, where $L=\{L_x,L_y\}$ is the linear dimension of the simulation box. The width, $L_x$, and height, $L_y$, of the simulation area are each set to each be integer multiples of the post spacing, such that the periodicity of the grid is continuous across the boundaries (see Table \ref{tab:MPCD_Parameters}).

To couple the sperm model to this solvent, the beads comprising the sperm's backbone are included in the collective collisions, allowing them to exchange momentum with the fluid~\cite{Malevanets2000}. 
Choosing a bead mass of $m_\text{s} = 10 m$, ensures a strong coupling of the sperm with the MPCD solvent. 
The beads are then included in the collision step (Eq. (\ref{eq:collision})) as solvent particles. In between two collision steps, the sperm backbone dynamics is integrated using a velocity-Verlet algorithm with a time step of $\tau_{MD}$\cite{Malevanets2000,Mussawisade2005} (see also Table \ref{tab:MPCD_Parameters}). While there are 
even stronger ways to couple a polymer to the MPCD~\cite{Theers2016a,Whitmer2010}, we found that this type of coupling strikes a good balance between coupling efficiency and computational cost. 

\subsection{Modeling of Post Arrays}

The same solvent coupling described in the previous section is also used to insert the posts into the simulations as boundaries. 
Each post consists of a set of beads constrained to set locations, with spacing $\delta$ between each bead along the periphery of the post.
While these beads cannot move, they are assigned a virtual mass ($m_\text{p} = 5 m$) and momentum, which is used during the collision step to influence the fluid. The virtual momentum is randomly assigned at each time step, sampled from a Gaussian distribution with zero mean and standard deviation $\sqrt{m k_B T}$, to keep it in thermal equilibrium with the fluid. 
Additionally, the beads forming the sperm interact with those forming the post using a truncated Lennard-Jones potential of the form~\cite{Yang2008}
\begin{equation}
E = \left\{\begin{array}{cc}
4 \epsilon \left[ \left(\frac{\sigma}{r}\right)^{12} - \left(\frac{\sigma}{r}\right)^6\right] & r \le \sqrt[6]{2}\sigma \\
-\epsilon & r > \sqrt[6]{2}\sigma \\
\end{array}\right.
\end{equation}
with range parameter, $\sigma$, and strength parameter, $\epsilon$.
These values are chosen with as small $\sigma$ and large $\epsilon$ as practical, to imitate a hard wall (see Table \ref{tab:MPCD_Parameters} for values),
preventing the sperm from crossing the posts. 

\begin{figure}[tb] 
\centering
\includegraphics[width=8.8cm]{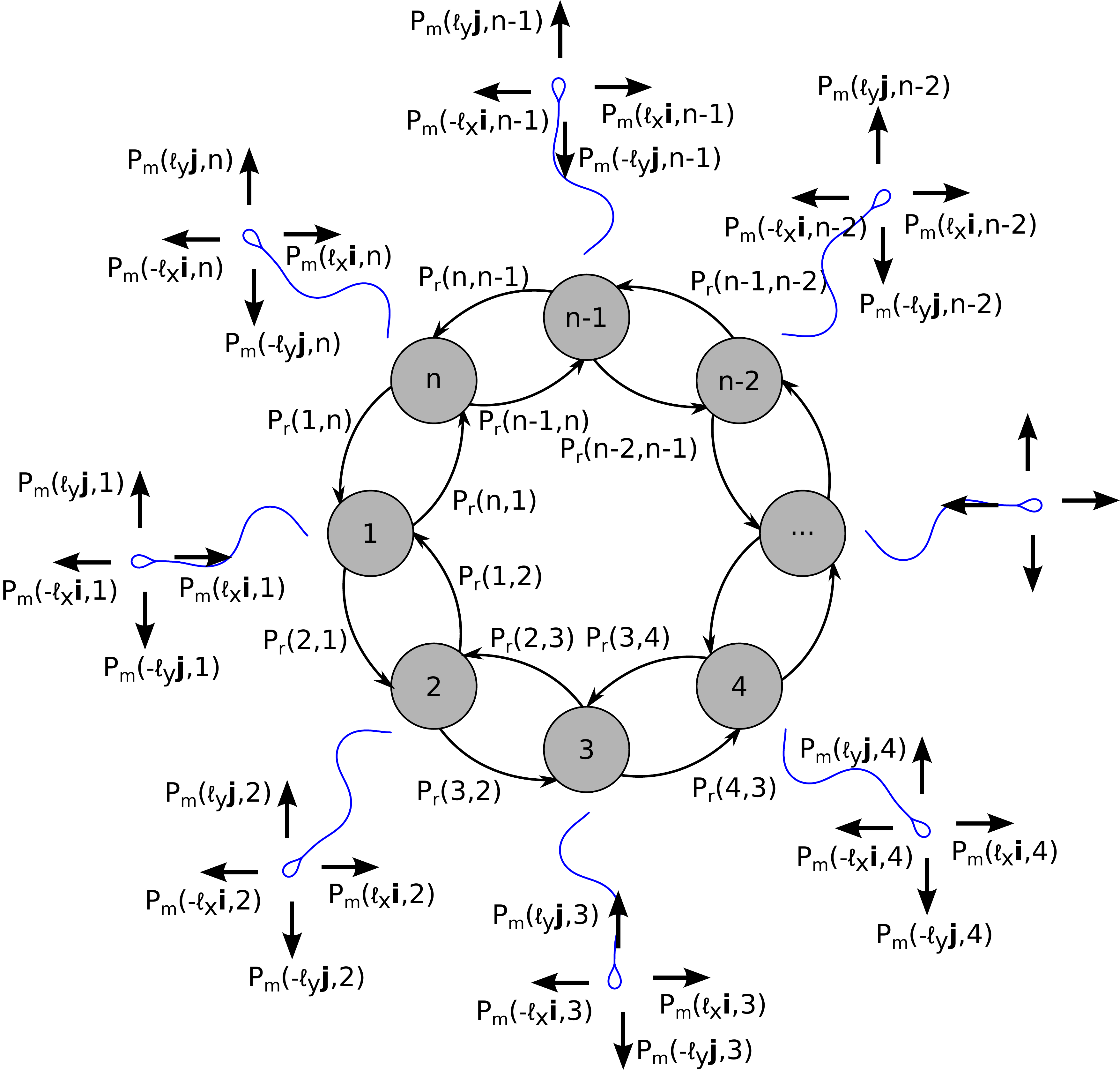}
\caption{State diagram of the stochastic lattice model.
At every time step, sperm transitions from a {\it current} rotational state $d$ to the {\it next} rotational state $d^\prime$ with the probability $P_r(d^\prime,d)$, as shown in the diagram.
Given a state $d$,  during the movement step of the simulation the sperm can move a distance of lattice spacing in one of the cardinal directions with the probability $P_m (\mv{\zeta}=\left\{0,\pm \ell_x \mv{i}, \pm \ell_y \mv{j}\right\}, d)$.  Note that  $P_r(d,d)$ corresponds to the probability of preserving the current rotational state.}
\label{fig:angle-markov}
\end{figure}

 \subsection{Thermostat}
In a closed system with a power source---the periodic forcing of the sperm's tail waveform for the case in question---the total energy of the system will increase with time.
To ensure that the overall temperature of the system is constant over time (and that the transport coefficients do not change), we employed the thermostat proposed by Huang et al.~\cite{Huang2010}.  The thermostat modifies the collision step (Eq. \ref{eq:collision}) as follows
\begin{equation}
\mv{v}_i(t+\tau) = \mv{u}_{c_i}+\lambda_{c_i} \mathcal{R}\left(\mv{v}_i-\mv{u}_{c_i}\right) \;\;,
\end{equation}
where 
\begin{equation}
\lambda_{c_i}=\sqrt{\frac{E^\prime}{E}}
\end{equation}
is a per-cell scale factor with the old collision energy defined as
\begin{equation}
E = \frac{1}{2} \sum_j m_j (\mv{v}_j-\mv{u}_{c_j})^2 \;\;.
\end{equation}
The new collision cell energy, $E^\prime$, is then chosen from a gamma distribution, i.e. $P(E^\prime)= \Gamma(N-1,k_BT)$ where 
$N$ is the number of particles in a collision cell at a given moment~\cite{Huang2010}.

\section{Lattice Model with Rotation}
\label{sec:lattice-model}

In order to model a large number of sperm cells moving over a long time period, as would be necessary to model a realistic device\cite{Chinnasamy2017}, we developed a probabilistic model for the movement of the sperm through the lattice formed by the post array.
This model takes into account that there are a finite number of directions that a sperm can be facing as it travels through the grid, and that its movement will depend on these  directions.
At every step, there is a probability that the sperm will turn left or right, changing its heading, i.e. switching to a different  rotational state, $d$.
As a function of its  current state, there is also a probability that the sperm will move in one of the four cardinal directions.
These probabilities can be calculated using the coarse-grained MPCD simulations (described in the previous section) for each condition that is considered. 
Mathematically, this process can be written as a Markov model with the two probabilities defined as
\begin{align}
P_m &= P_m(\mv{\zeta},d)  \;\;, \\
P_r &= P_r(d^\prime, d) \;\;, 
\end{align}
where $P_m$ and  $P_r$ are the {\it movement} and {\it rotation} probabilities, respectively. Here, 
$\mv{\zeta}=\left\{0,\pm \ell_x \mv{i}, \pm \ell_y \mv{j}\right\}$ corresponds to the 
spatial increment in lattice coordinates,  $d$ and $d^\prime$ correspond to the current  and next rotational states, respectively.
These probabilities and the transitions between each state are illustrated in the state diagram in Fig. \ref{fig:angle-markov}. 

In the limit of many sperm cells, we can consider the sperm population as a density, $\rho(\tilde{\mv{R}},d,t)$, where $\tilde{\mv{R}}\equiv \xi \ell_x\mv{i} +  \chi \ell_y \mv{j}$ is the position vector in lattice coordinates with $\xi,\chi$ integers, and, $\{\mv{i},\mv{j}\}$ are the cardinal direction unit vectors.
The time evolution of this density function then can be written as movement and rotation steps performed during one time step, and this time 
step is chosen to be the period of sperm tail oscillation, $T$. Note that this choice simplifies the analysis of sperm motility by stroboscopically eliminating sperm
head wobble. This density time evolution can be represented in two half-time steps ($T/2$) given by

\begin{eqnarray}
 \rho (\tilde{\mv{R}},d,t-T/2) &=& \sum_\mv{\zeta}P_m(\mv{\zeta},d)\rho(\tilde{\mv{R}}-\mv{\zeta},d,t-T) \nonumber \;\;, \\
 \rho (\tilde{\mv{R}},d^\prime,t) &=& \sum_{d}P_r(d^\prime,d)\rho(\tilde{\mv{R}},d,t-T/2) \;.
\end{eqnarray}

\begin{figure}[htb] 
\centering
\includegraphics[width=8.6cm]{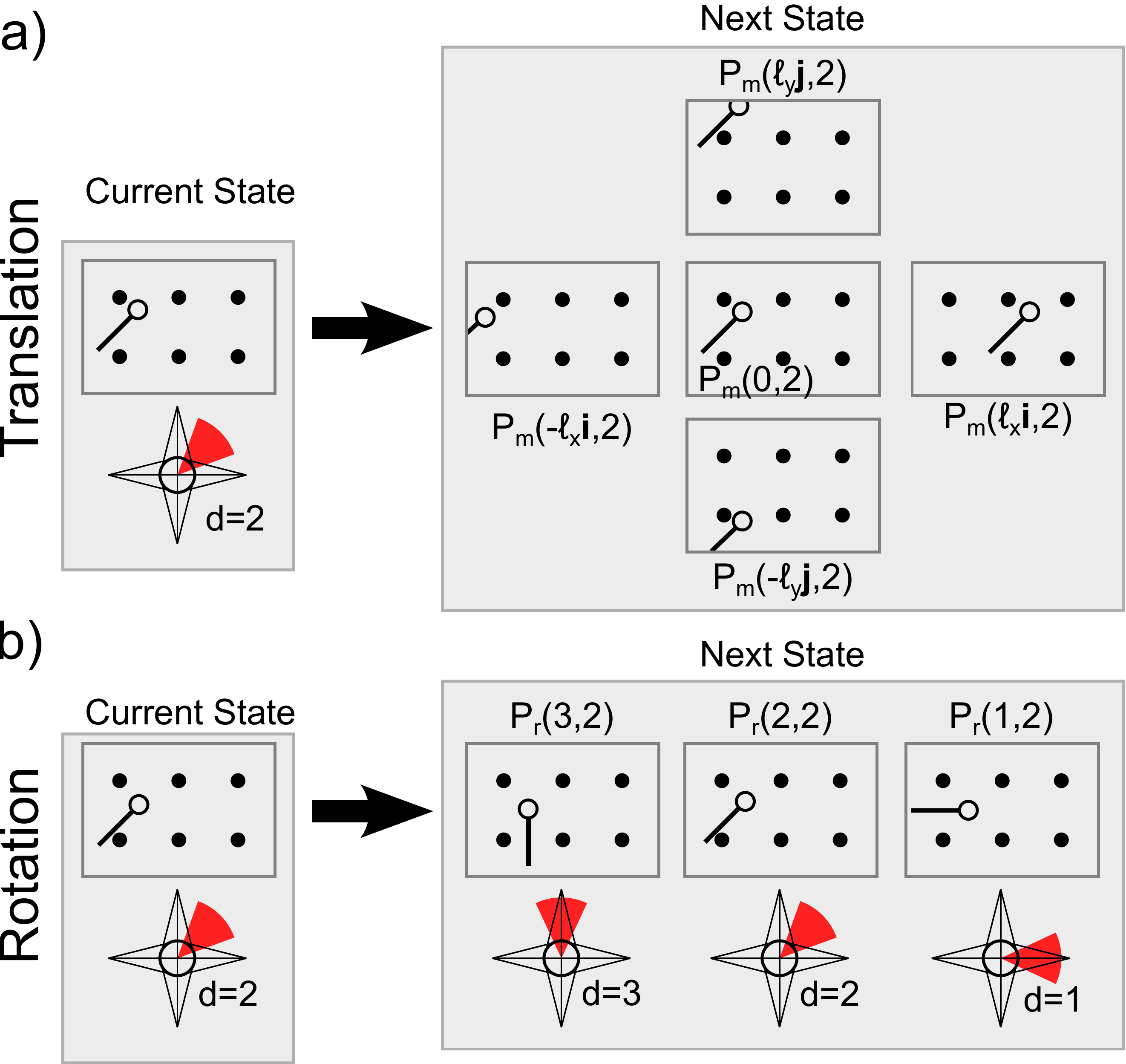}
\caption{Illustration of a single time step of the stochastic lattice model.
At every such step, the sperm first takes a translational move in one of the four cardinal directions (regardless of its orientation), based on the probability, $P_m$, for its current direction $d$.
The sperm then changes its rotational state $d$ to a new direction state $d^\prime$ based on the transition probability, $P_r$, as illustrated in  Figure \ref{fig:angle-markov}.
a) For example, for a sperm with rotational state $d=2$, the probability of taking a step to the left is $P_m(-\ell_x \textbf{i},2)$, and the probability of taking a step downwards is $P_m(-\ell_y \textbf{j},2)$.
b) After choosing a new position as directed by the movement probabilities, the sperm will have the option to transition into a different rotational state, either $d=1$ or $d=3$ with the transition probabilities $P_r(1,2)$ and $P_r(3,2)$, respectively.
The probabilities $P_m$ and $P_r$ are determined using the MPCD simulations.}
\label{fig:angle-markov-example}
\end{figure}

This two-step dynamics is illustrated in Fig. \ref{fig:angle-markov-example}. First, consider a sperm in a given state, which we will refer to as the {\it current state}. We then
calculate the probabilities of this sperm cell moving one lattice site along one of the cardinal directions or staying put (Fig. \ref{fig:angle-markov-example}a), i.e. transitioning to the {\it next state}, while preserving its current rotational state ($d=2$ in Fig. \ref{fig:angle-markov-example}a). 
In the second sub-step, once again given the current state, the probability of transitioning to one of the adjacent rotational states is calculated. In the example in Fig. \ref{fig:angle-markov-example}b, the sperm can transition into either direction state $d=1$ or $d=3$.  These two steps 
 are then repeated at each time-step for a given sperm.

It is also useful to define a population density that is  summed over rotational states at a given location, namely

\begin{equation}
 \rho_d(\tilde{\mv{R}},t) = \sum_d\rho(\tilde{\mv{R}},d,t) \;\;.
\end{equation}
Here, $ \rho_d(\tilde{\mv{R}},t)$ corresponds to the density of observable sperm in a given location, and can be used to calculate spatial density profiles of cells (see Ref. ~\cite{Chinnasamy2017}). 

\begin{figure}[htb] 
\centering
\includegraphics[width=8.6cm]{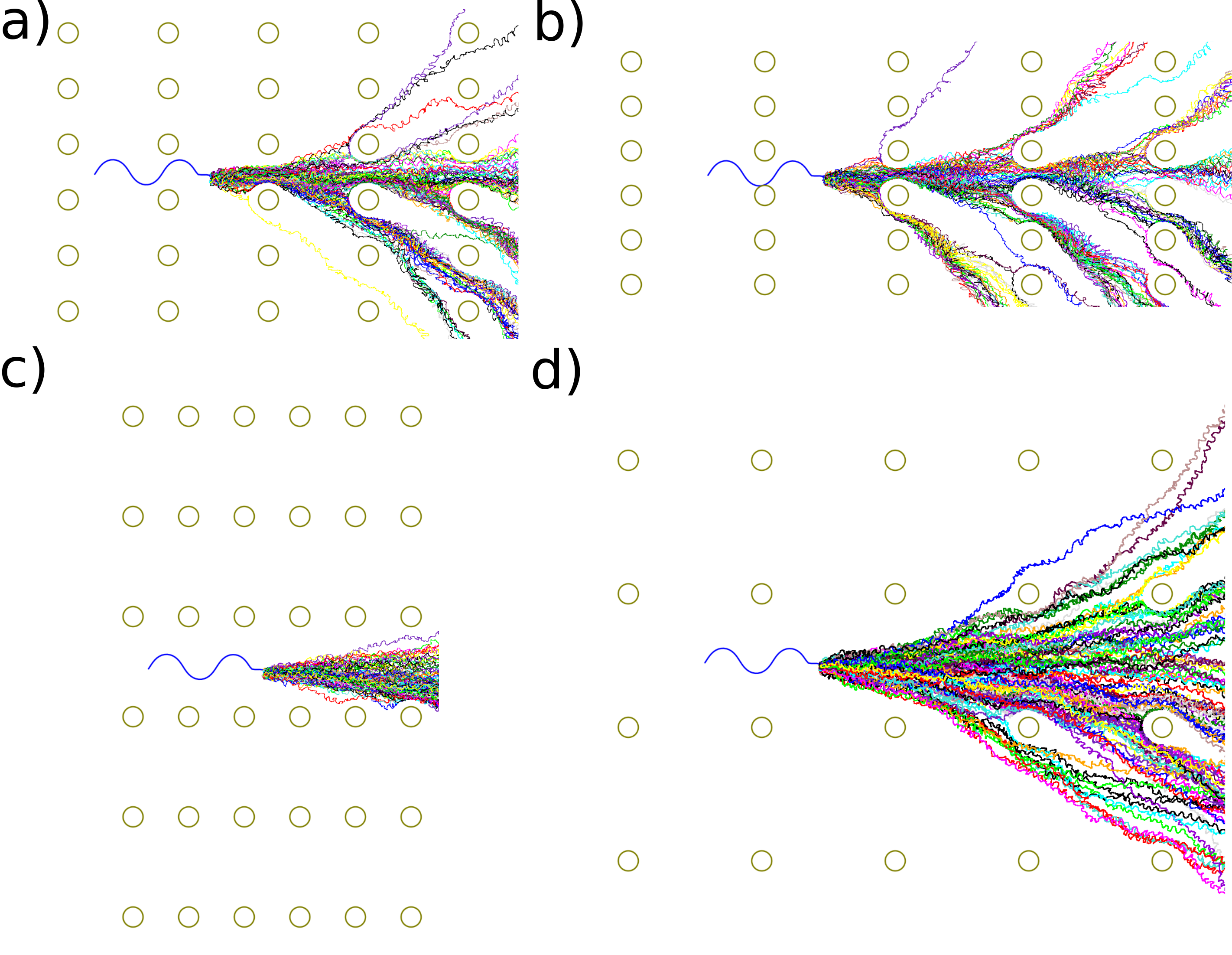}
\caption{Example sperm trajectories for 100 sperm in post spacings of (a) $\{\ell_x,\ell_y\}=36\times 20$, (b) $48\times 16$, (c) $20\times 36$, and (d) $48\times 48$,
in units of collision cell size $a$. $100$ trajectories for each spacing is shown. }
\label{fig:SpermTraceExamples}
\end{figure}

\begin{figure}[htb]. 
\centering
\includegraphics[width=8.6cm]{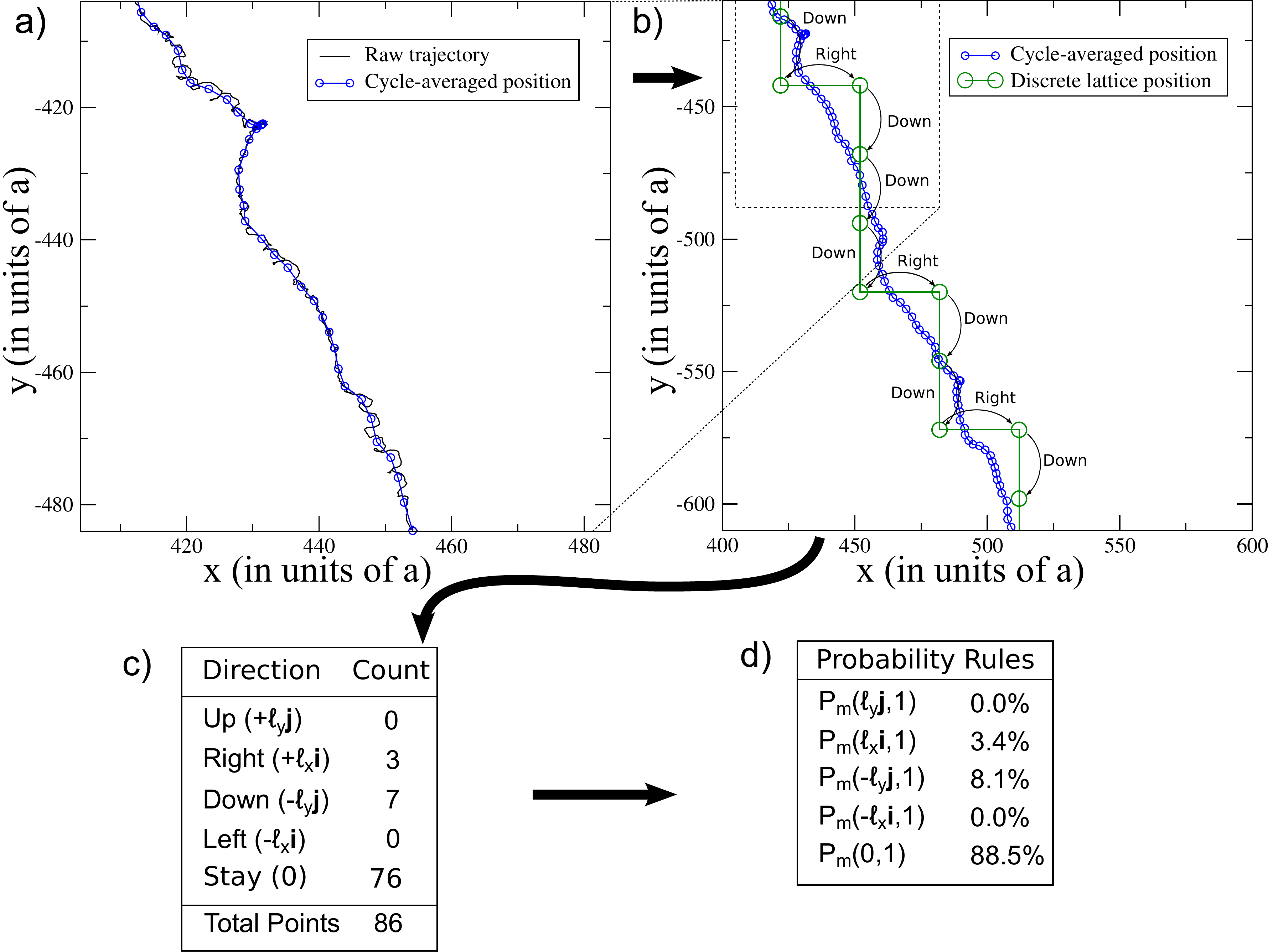}
\caption{
Example walkthrough of the conversion from raw MPCD trajectory data to probabilities for movement.
a) The initial MPCD position data (black lines) is averaged such that there is one point per tail cycle (blue circles).
b) The MPCD position data is then discretized into integer lattice positions.
c) Transitions between lattice positions are recorded and summed.
d) These counts of transitions are divided by the total number of tail cycles to form movement probabilities.
}
\label{fig:srd-to-lattice-position}
\end{figure}

\begin{figure}[htb]  
\centering
\includegraphics[width=8.6cm]{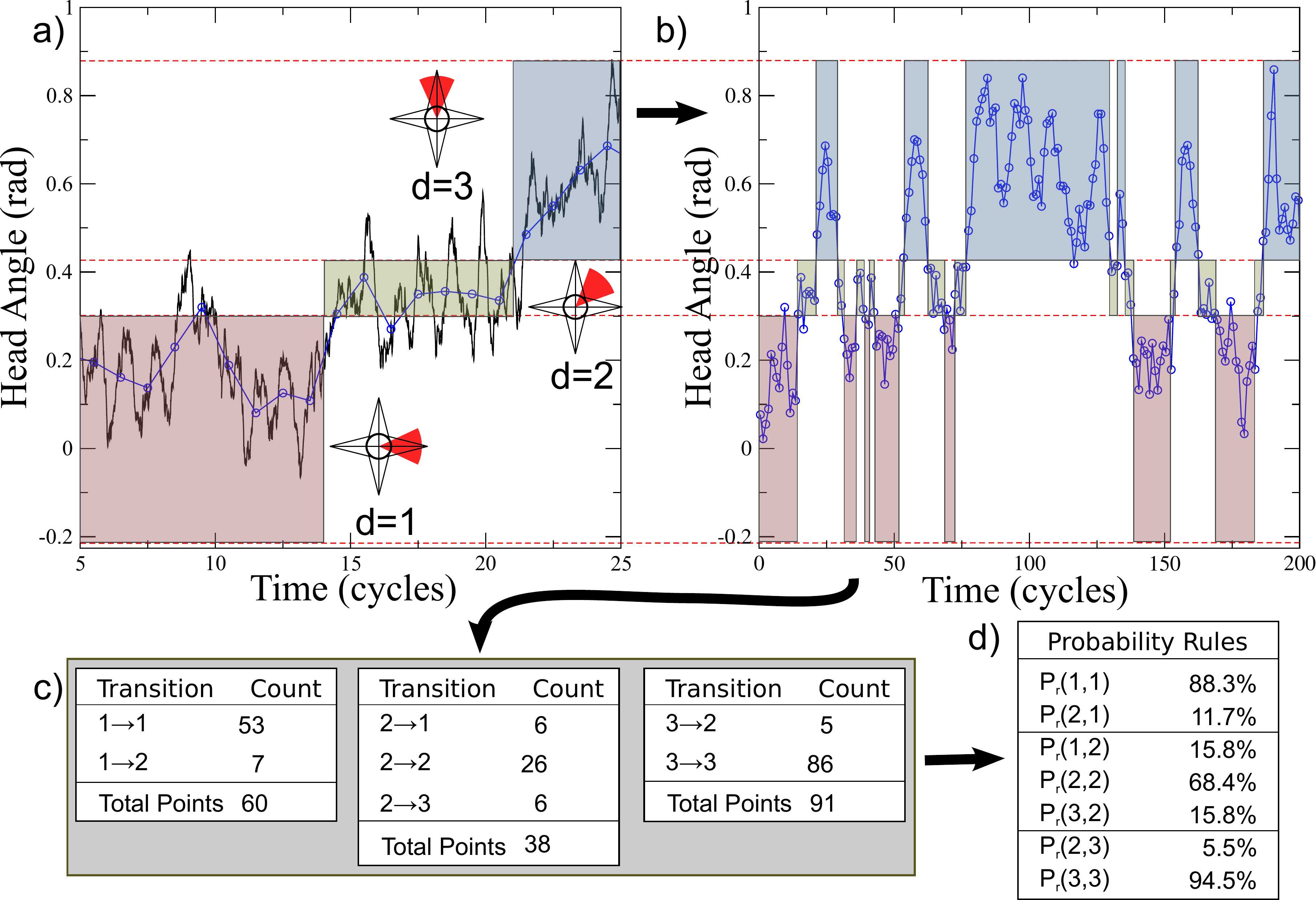}
\caption{
Example walkthrough of the conversion from raw MPCD trajectory data to rotational probabilities.
a) The initial MPCD angle data (black lines) is averaged such that there is one point per tail cycle (blue circles).
b) The  MPCD angle data is then separated into categories.
c) Transitions between categories are recorded and summed.
d) These counts of transitions are divided by the total number of tail cycles per category to form rotational probabilities.
}
\label{fig:srd-to-lattice-angle}
\end{figure}

\subsection{Calculation of probabilities}
\label{sec:model-processing}

For each configuration under examination, the initial data is collected by running $100$ independent iterations of the MPCD simulation, with that configuration.
Each one of these simulations produces a trajectory of the sperm (as defined by the center of mass of the head), as well as its angle (defined as the vector connecting the center of mass of the head and the attachment point of the neck), as a function of time.
Sample trajectories for four different spacings illustrating the diverse set of behaviors of sperm in the post arrays are shown in Fig. \ref{fig:SpermTraceExamples}.
In order to limit the amount of computation required to determine movement probabilities, the symmetry of 
the system is used to count every trajectory both as itself and as if it was rotated 180 degrees. 
In other words, the trajectory in which the sperm starts out facing left, and then goes up is equivalent to the one in which it starts out facing right and then goes down.

The lattice model does not directly use these trajectories, however, it needs the transition probabilities for the sperm moving and turning at each step. 
Therefore, a conversion from trajectory data in MPCD units ($a$ and $\tau$) to the lattice model coordinates which requires units of post-spacing ($\ell_x$,$\ell_y$) and sperm tail-cycles ($T$). 
To accomplish this, the positional and directional data is first block-averaged down to one point per swimming cycle (Figs. \ref{fig:srd-to-lattice-position}a and \ref{fig:srd-to-lattice-angle}a).
From here, the position data is then coarse-grained into integer coordinates on a grid defined by the posts (Figure \ref{fig:srd-to-lattice-position}b).
Positional noise and sampling error is suppressed by removing any single-point discontinuities in the discretized position and angle.
This prevents a single transition from appearing to be interpreted as multiple transitions, due to noise and fluctuations across the dividing line between two states.

To categorize the different directions that the sperm can face, a histogram of the distribution of angles is created, and its minima are used to define directional categories (shaded areas in Figs. \ref{fig:srd-to-lattice-angle}a and \ref{fig:srd-to-lattice-angle}b).
Each category here forms a discrete value of $d$ that the sperm can adopt as a direction.
The position data is then sorted into segments based on their directional categories.
For all the segments in each category, the total number of steps along the lattice, in each direction is counted up (Figure \ref{fig:srd-to-lattice-position}c).
These totals are then divided by the number of swimming cycles that make up those segments, to produce final movement probabilities for that category (Figure \ref{fig:srd-to-lattice-position}d).
Similarly, the direction data is split into categories (Figs. \ref{fig:srd-to-lattice-angle}a and \ref{fig:srd-to-lattice-angle}b), and the total number of transitions between these categories 
is counted (Fig. \ref{fig:srd-to-lattice-angle}c). 
These transitions are then divided by the number of cycles spent in each category, to again produce a set of rotational probabilities (Figure \ref{fig:srd-to-lattice-angle}d).

\begin{figure} 
\centering
\includegraphics[width=8.6cm]{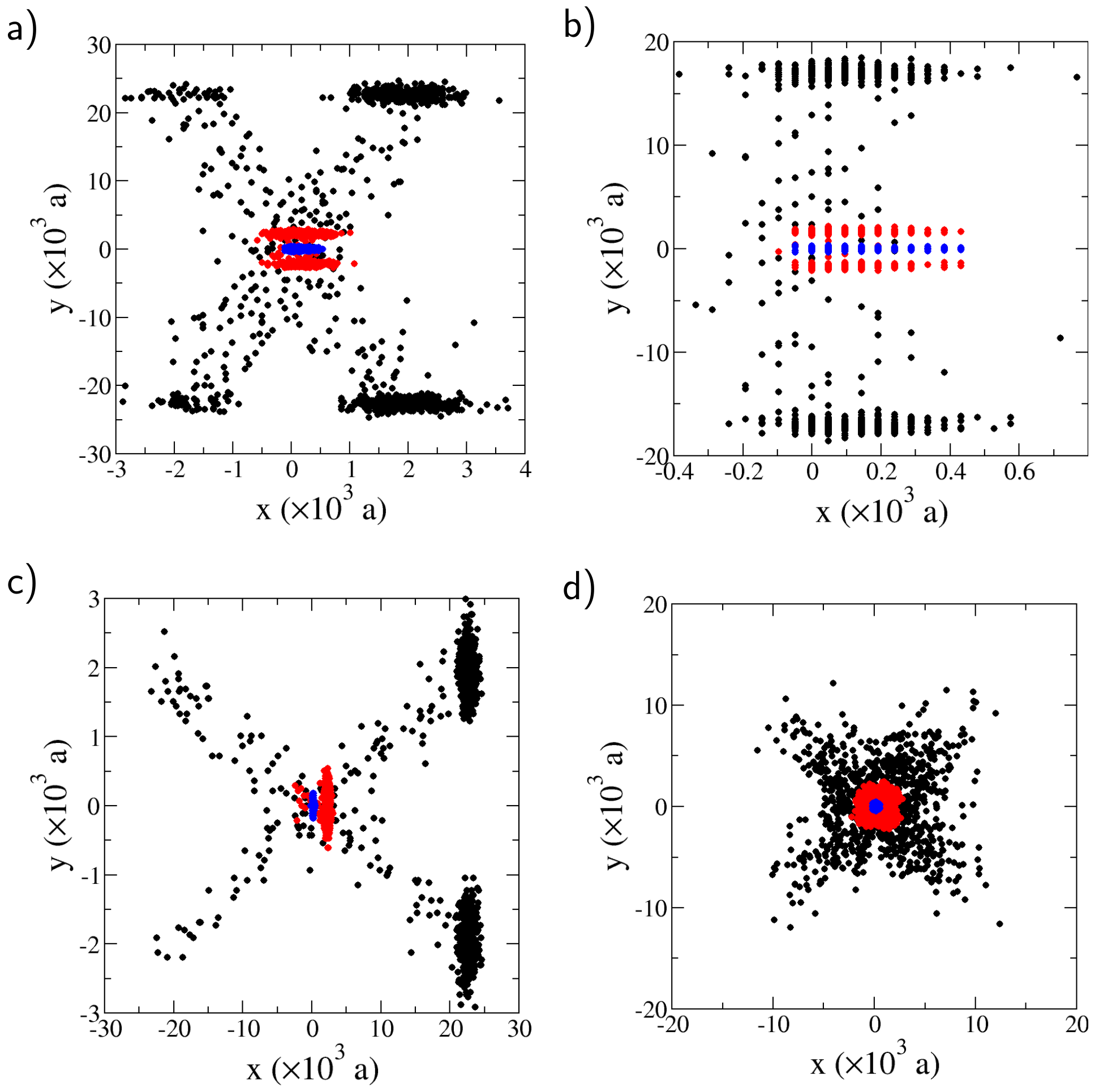}
\caption{Distribution of 1000 sperm as produced from the lattice model, for the trajectories in Figure \ref{fig:SpermTraceExamples}.
Point clouds are presented for points at $\Delta t/T=10^2$ (blue), $\Delta t/T=10^3$ (red), and $\Delta t/T=10^4$ (black), where $T$ is the length of a cycle, i.e. $T=2\pi/\omega$.
Post spacings are (a) $\{\ell_x,\ell_y\}=36\times 20$, (b) $48\times 16$, (c) $20\times 36$, and (d) $48\times 48$, in units of $a$. 
}
\label{fig:lattice-distribution}
\end{figure}

To illustrate the lattice model's versatility at long length and time scales, the four post spacings used in Fig. \ref{fig:SpermTraceExamples} were used to generate trajectories of 1000 sperm for a duration $\Delta t/T=10^{4}$; results are shown in Fig. \ref{fig:lattice-distribution}.  
Such long time behavior further amplifies the subtle differences in motility shown in Fig.  \ref{fig:SpermTraceExamples},  
resulting in different localization of sperm cell populations at large distances and long times (see Fig. \ref{fig:lattice-distribution}), which has direct consequences for device efficiency, i.e. sperm collection~\cite{Chinnasamy2017}. 
It is also important to note that the simulation of all of these trajectories takes on the order of a several seconds, in contrast to an equivalent calculation with the MPCD model in which a single sperm trajectory of the same duration would take several weeks. 
    
\section{Results} \label{sec:results}

\subsection{Enhancement of sperm velocity and persistence}

Standard practice in the field of fertility~\cite{Gardner2013}, is to characterize sperm trajectories via either {\it straight-line velocity} (VSL), {\it curvilinear velocity} (VCL), or {\it average path velocity} (VAP) ~\cite{Amann2014,Tasoglu2013}.
While these velocities give useful and complementary information about sperm motility behavior, they have dependence on the experimental conditions used to gather them.

For instance, an accurate value of VCL---defined based on the total path-length traveled by the sperm head---requires a frame rate significantly higher than the rate of the sperm head wobble to avoid aliasing artifacts.
On the other hand, conventional calculation of VAP is performed by a running average of the position of the sperm's head, with a constant number of frames used for that average.
This means that as frame rate increases, the value of VAP approaches VCL, as the time represented by the average decreases.
A contrasting problem exists for VSL, which is based on a straight line drawn from the first to last position of the sperm, introducing a dependence on the length of time observed.

To avoid these pitfalls, here we will analyze sperm trajectories from our simulations by calculating the Mean-Squared-Displacement (MSD) in two dimensions using 

\begin{equation}
\MSD = \langle (x(t)-x(0))^2 \rangle + \langle (y(t)-y(0))^2 \rangle \;\;. 
\end{equation}

In our earlier work~\cite{Tasoglu2013}, we demonstrated that sperm motility in blank microfluidic channels can be described by the Persistent Random Walk (PRW) model\cite{Dickinson1993,Chinnasamy2017}. In this model, the MSD takes the form

\begin{equation}
\label{eq:PRW-MSD}
\MSD = 2 L_p\left[ St-L_p\left(1-e^{St/L_p}\right)\right] \;\;,
 \end{equation}
\noindent where $S$ corresponds to short-time velocity of the sperm, and $L_p$ denotes its characteristic persistence length.
One can easily show that in this model, the motion of sperm is ballistic at short times ($\MSD \sim t^2$), and diffusive at long times ($\MSD \sim t$).
We then fit the data generated using our MPCD model to PRW model to calculate $S$ and $L_p$ values for each of the conditions simulated.

\begin{figure} 
\centering
\includegraphics[width=8.6cm]{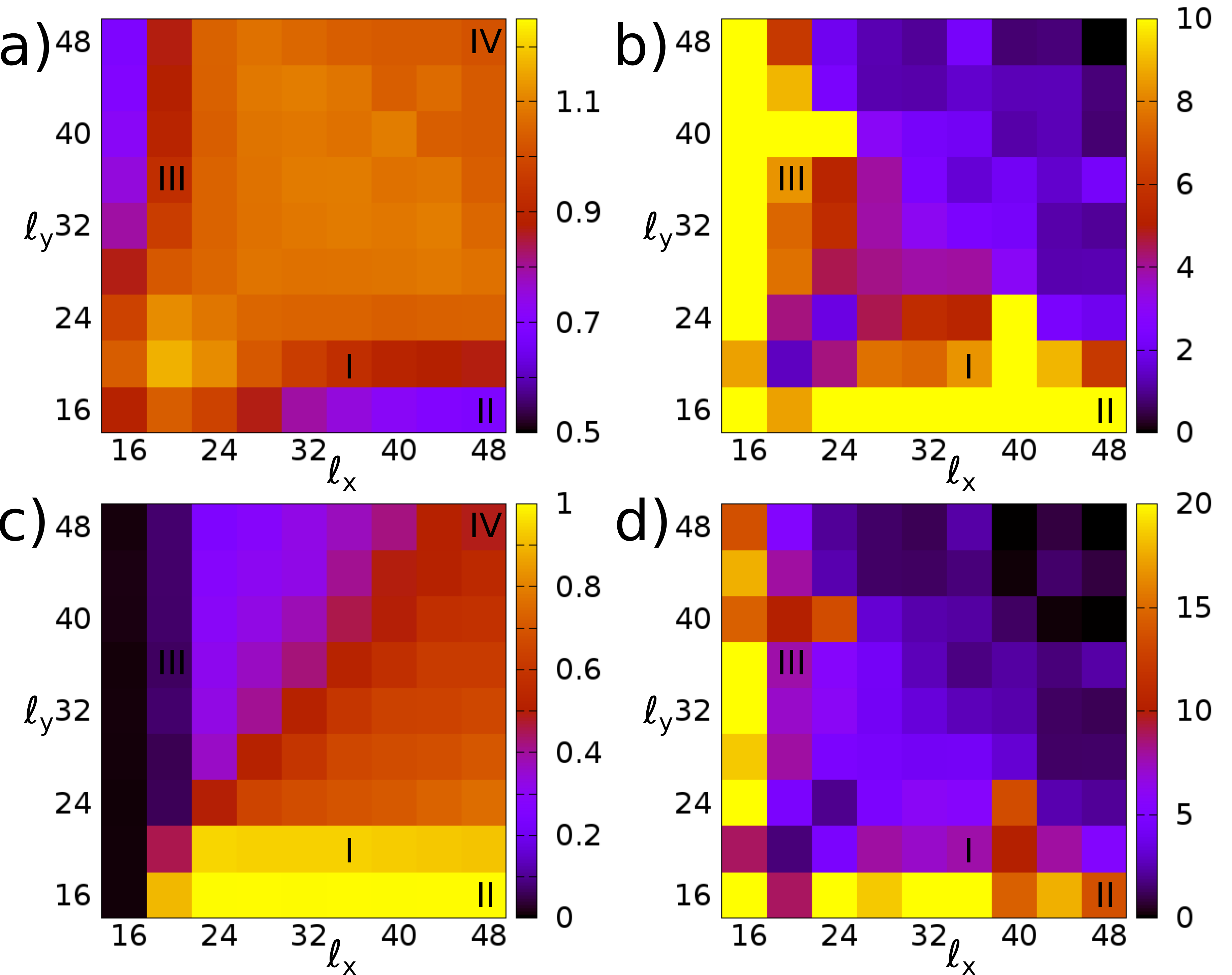}
\caption{Efficiency of different post spacings illustrated via heat maps. (a) Normalized  speed $S/S_0$, 
(b) normalized persistence length $L_p/L_p^0$, (c) propensity $\gamma$, and (d) effective diffusion coefficient, $D_{\text{eff}}$, 
are shown for varying spacings.  $S_0$ and $L_p^0$ correspond to the speed and persistence length of 
an unconstrained sperm in an infinite system, respectively. }
\label{fig:MSD-heatmaps}
\end{figure}

Normalized speed, $S$, and persistence length, $L_p$, values obtained from these fits are shown in Fig. \ref{fig:MSD-heatmaps}a and b.
The most speed enhancement is observed for the $20 \times 20$ post spacing in contrast to a nominal change in persistence length at this spacing.
This speed up is due to hydrodynamic interactions between the sperm and the posts.
It is also interesting to note that the $48\times16$ (labeled II in Fig. \ref{fig:MSD-heatmaps}) is an extreme at which 
the speed is lowered due to close confinement without benefiting hydrodynamically from the posts.
This close confinement also gives an extremely high value of $L_p$.

Post spacings with nearly a factor of two of each other are located in the two regions of high $L_p$ at the asymmetric edges.
For instance, spacings of $36\times 20$ (labeled I), and $20\times 36$ (labeled III) result in high persistence and speed, making them ideal for increasing the distances sperm cover in the array, and making them efficient transporters.
Finally, at the extreme of the post spacings we considered, namely 
$48\times 48$ (labeled IV), we see little to no benefit from the posts.
At such spacings, the posts merely become an occasional impediment to transport.

\subsection{Propensity of sperm}
In order to calculate propensity of sperm, we utilize the lattice model, as it directly encodes information about the behavior of the sperm at long times.
To do this, we first define a direction average of sperm density, namely, 

\begin{equation}
 \rho_{\tilde{\mv{R}}}(d,t) = \sum_{\tilde{\mv{R}}}\rho(\tilde{\mv{R}},d,t) \;\;.
\end{equation}

At steady state $\rho_{\tilde{\mv{R}}}(d,t_{\text{s}}) - \rho_{\tilde{\mv{R}}}(d,t_{\text{s}}+T) \rightarrow 0$, hence one can write 
\begin{eqnarray}
 &\rho_{\tilde{\mv{R}}}(d,t_{\text{s}})\!=\! \rho_{\tilde{\mv{R}}}(d,t_{\text{s}})\left[1\!-\!P_r(d\!+\!1,d)\!+\!P_r(d\!-\!1,d)\right]  \nonumber \\ 
 & \!+\!\rho_{\tilde{\mv{R}}}(d\!-\!1,t_{\text{s}})P_r(d,d\!-\!1)  \nonumber \\
 & \!+\!\rho_{\tilde{\mv{R}}}(d\!+\!1,t_{\text{s}})P_r(d,d\!+\!1) \;\;,
\end{eqnarray}
\noindent where $t_{\text{s}}$ is denotes an arbitrary time point at steady state.  As the probability rules are perfectly symmetric, we want to count all vertical moves together and all horizontal moves together. Density weighted average probability of moving
in a given direction $\mv{\zeta}$ is then given by

\begin{equation}
\gamma_{\mv{\zeta}} = \frac{\sum_d \rho_{\tilde{\mv{R}}}(d)P_m(\mv{\zeta},d) }{\sum_d \rho_{\tilde{\mv{R}}}(d)} \;\;,
\end{equation}
\noindent where we dropped time dependence of density for brevity. To produce a parameter quantifying directional persistence, we break the symmetry of our probability rules 
and consider movement only in the positive cardinal directions. This can be formally done by defining the {\it propensity} of sperm, $\gamma$, as

\begin{equation}
\gamma = \frac{2}{\pi} \arctan \left(\frac{\gamma_{\mv{j}} \ell_y}{\gamma_{\mv{i}} \ell_x}\right) \;\;. 
\end{equation}

Note that $\gamma$ is in the range $[0,1)$, which quantifies the tendency for the sperm to travel in the horizontal or vertical directions of the grid.
This is related to, but independent from the persistence length, $L_p$, because while an extreme value of $\gamma$ will correspond to a large $L_p$, a medium value of $\gamma$ could either correspond to rapid changes in direction (low $L_p$), or continued persistence along the diagonal of the grid. 

\subsection{Effective diffusion and phase diagram}

One can show that in the long time limit Eq. (\ref{eq:PRW-MSD}) reduces to $\MSD \approx 2S L_pt\equiv 4 D_\text{eff} t$, where 
$D_\text{eff} = SL_p/2$. Fig. \ref{fig:MSD-heatmaps}d shows this effective diffusion coefficient for different post spacings, illustrating the combined effect of speed and persistence length.
Our results show that the most enhancement of diffusion occurs for spacings that have about $50\%$ difference in aspect ratios. 
It is also important to note that points with minimal $D_\text{eff}$ enhancement are primarily found with medium size aspect ratios, which results in minimal directionality preference, i.e. $\gamma \approx 0.5$.
This is in contrast to the point with a high directionality preference---$\gamma \approx 0$ and $\gamma \approx 1$---which exhibit a strong enhancement in effective diffusion coefficient due to that high $L_p$. 

The effects of a strong propensity can be clearly seen in trajectories at long times, as shown earlier in \figformat{Fig. \ref{fig:lattice-distribution}}. For instance, 
\figformat{Fig. \ref{fig:lattice-distribution}a} and \figformat{c} show a factor of ten directional preference for sperm to move along $x$ or $y$ axes, and \figformat{Fig. \ref{fig:lattice-distribution}b} shows very tight constraint, reorienting all of the sperm from moving along $x$ to along $y$  within ten post spacings.
In contrast, \figformat{Fig. \ref{fig:lattice-distribution}d} shows a symmetric diagonal-preferring behavior for sperm, covering less than half the distance of the 
configurations with higher persistence length.

\begin{figure} 
\centering
\includegraphics[width=8.4cm]{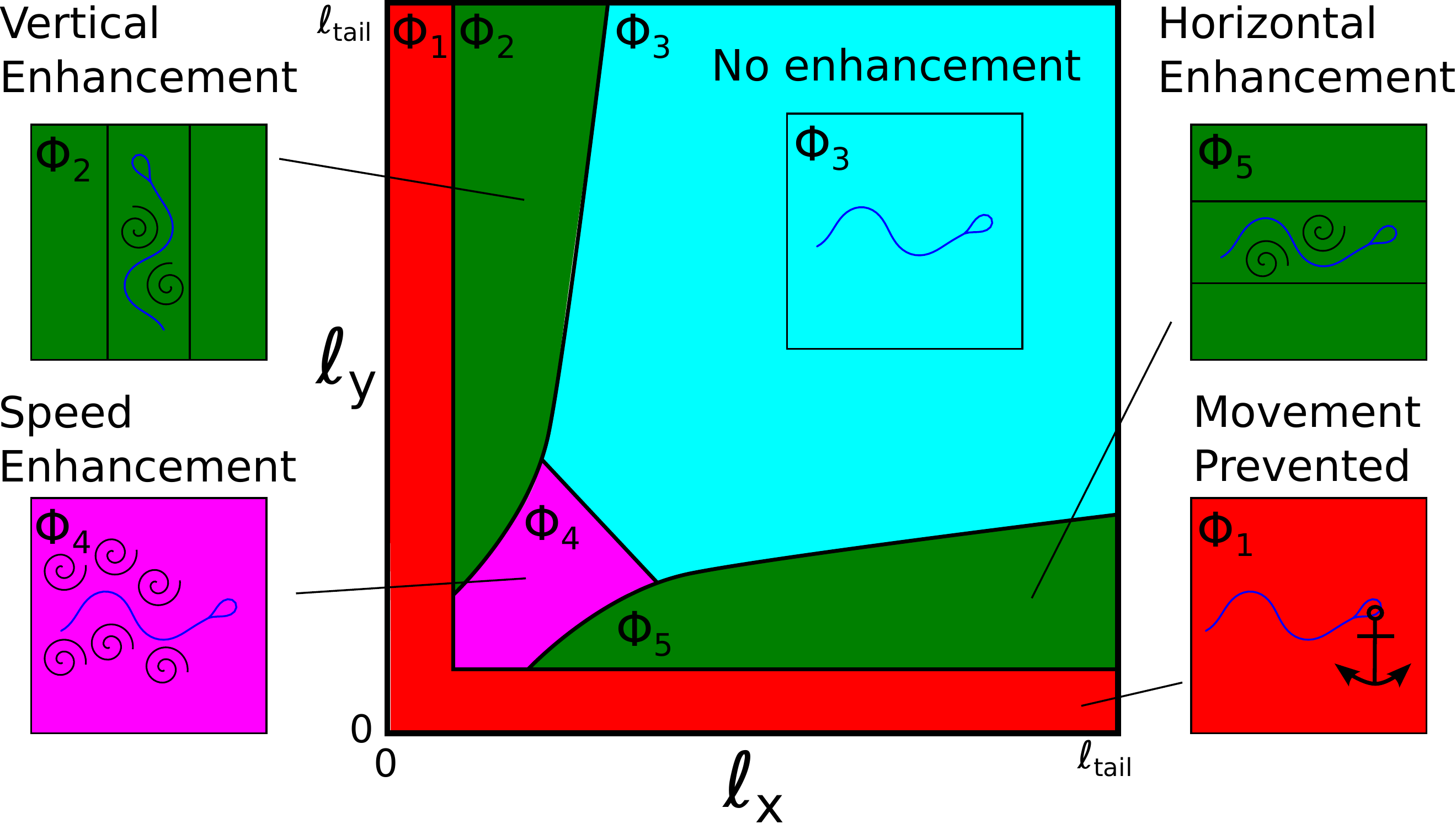}
\caption{Phase diagram of sperm motility in post arrays. }
\label{fig:phaseDiagram}
\end{figure}

All of our results are summarized in the phase diagram shown in Fig. \ref{fig:phaseDiagram}. Region $\Phi_1$ corresponds to no movement zone where sperm is practically anchored. Here, regions $\Phi_2$ and $\Phi_5$ are the horizontal and vertical confinement zones. While no enhancement is observed in region $\Phi_3$, Region $\Phi_4$ shows the most enhancement in speed. 

\section{Conclusions}

In this paper, we explored how sperm behave in a periodic array of posts, motivated by a recently introduced microfluidic device for sperm sorting for infertility treatment~\cite{Chinnasamy2017}.
Our approach uses a novel multi-scale model, and it combines the strength of a particle-based mesoscale solvent with a probabilistic model. 
Like all pusher swimmers, sperm are hydrodynamically attracted to a surface, eventually causing them to follow, even turn corners~\cite{Eamer2016}. 
However, when the surface has gaps of sufficient size and a favorable geometry, the sperm will turn towards and swim through such a gap, resulting in a net steering of sperm in a different direction. 
It is therefore expected to find a wide variety of swimming behaviors due this competition between hydrodynamics and boundary effects in post array geometries. 

We indeed found a rich phase diagram of various transport 
modalities, while the most speed enhancement ($\sim 20\%$) was found for a symmetric spacing, increased persistence was observed for asymmetric 
arrangement of posts ($\sim 50\%$ difference in aspect ratio). 
This effect, at first glance is somewhat paradoxical, as one would naively expect 
that sperm would take the path of least resistance---i.e. wider openings. However, when the posts are fairly close together, they behave like a porous wall, and the sperm are attracted to them. When properly placed, our results show that asymmetry pays off and results in enhanced flagellar transport. 

Our results show that any microfluidic device that has the goal of sorting sperm needs to 
strike a balance between enhanced speed and increased persistence length, while maintaining a propensity to move towards the outlet, 
 in order to maximally separate healthy sperm with good swimming behavior from unhealthy sperm which swim poorly, or simply are diffusive.
More specifically, our phase diagram suggests that regions $\Phi_2$ and $\Phi_5$ would be most suited for a sperm sorting using a post array device, and this conclusion is consistent with our recent experimental observations~\cite{Chinnasamy2017}. 

Our model can also be applied to sperm cells with morphological defects, and help explain how such a device works, essentially like a filter leaving behind sperm with motility or shape defects~\cite{Chinnasamy2017}. 

We believe our findings will have implications not only in reproductive medicine, but also in bio-hybrid robotics that use scaled-down actuation systems that mimic flagellar swimmers for applications in bio-diagnostics. 

\begin{acknowledgments}
We thank all the members of T\" uzel group and Demirci Lab for helpful discussions, particularly Steven Vandal for insightful suggestions on the analysis of the large parameter space.
We would like to particularly thank Dr. Christopher R. Lambert (Worcester Polytechnic Institute, WPI), Dr. Fatih \. Inci (Stanford) and Dr. Thiruppathiraja Chinnasamy (Stanford) for their insightful suggestions.
This work was supported by NSF-CBET 118399, NSF-CBET 1309933, and NIH R01EB015776.
ET also acknowledges support from WPI Startup Funds.
JLK acknowledges support from the WPI Alden Fellowship. 
\end{acknowledgments}

\end{document}